\documentclass[epj,widetext]{svjour}
\usepackage{graphicx}
\usepackage{mathptmx, textcomp}
\usepackage[latin1]{inputenc}
\usepackage{braket}
\usepackage{array}
\usepackage{float}

\begin{document}

\title{Production of a dual-species Bose-Einstein condensate of Rb and Cs atoms}
\author{A. D. Lercher\inst{1}
\and T. Takekoshi\inst{1}
\and M. Debatin\inst{1}
\and B. Schuster\inst{1}
\and R. Rameshan\inst{1}
\and F. Ferlaino\inst{1}
\and R. Grimm\inst{1,2}
\and H.-C. N\"{a}gerl\inst{1}}
\institute{
Institut f\"ur Experimentalphysik, Universit\"at Innsbruck,
6020 Innsbruck, Austria
\and
Institut f\"ur Quantenoptik und Quanteninformation,
\"Osterreichische Akademie der Wissenschaften, 6020 Innsbruck,
Austria
}
\date{Received: 4 January 2011}
\abstract{
We report the simultaneous production of Bose-Einstein condensates (BECs) of $^{87}$Rb and $^{133}$Cs atoms in separate optical traps. The two samples are mixed during laser cooling and loading but are separated by $400~\mu$m for the final stage of evaporative cooling.  This is done to avoid considerable interspecies three-body recombination, which causes heating and evaporative loss. We characterize the BEC production process, discuss limitations, and outline the use of the dual-species BEC in future experiments to produce rovibronic ground state molecules, including a scheme facilitated by the superfluid-to-Mott-insulator (SF-MI) phase transition.
}

\maketitle

\section{Introduction}
\label{intro}
Over the last 10 years, quantum gas mixtures have been the focus of intense experimental and theoretical interest.  Degenerate Bose-Fermi \cite{Hadzibabic2002tsm,Roati2002fbq,Inouye2004ooh}, Bose-Bose \cite{Modugno2002tas}, and Fermi-Fermi \cite{Taglieber2008qdt,Spiegelhalder2009cso} mixtures have greatly enriched the spectrum of possible experiments. In particular, mixed quantum gases have recently attracted widespread attention as a starting point for efficient creation of high phase-space density samples of weakly bound heteronuclear Feshbach molecules and, after molecular state transfer, the creation of ultracold and dense samples of dipolar ground-state molecules \cite{Ni2008ahp}. Such samples promise myriad possibilities for new research directions in fields such as ultracold chemistry, quantum information science, precision metrology, and Bose-Einstein condensation \cite{Krems2008ccc,Demille2002qcw,DeMille2008est,Zelevinsky2008pto,Goral2002qpo,Damski2003coa,Baranov2008tpi,Pupillo2009cmt,Lahaye2009tpo,Carr2009cau,Friedrich2009wac}.

Heteronuclear molecules such as KRb and RbCs in their rovibronic ground state feature a sizable permanent electric dipole moment of about 1 Debye \cite{Kotochigova2005air,Deiglmayr2008cos}, giving rise to long-range and anisotropic dipole-dipole interactions when subjected to an external polarizing electric field. As a consequence, at ultralow temperatures, scattering and reaction dynamics are expected to be dominated by dipolar effects \cite{Micheli2007cpm,Li2008uic,Quemener2010sdo,Quemener2010efs,Idziaszek2010urc,Idziaszek2010sqm,Micheli2010urf}. Recent experiments with KRb have in fact reached this regime \cite{Ni2010dco}. A broad variety of novel quantum phases have been proposed for bosonic and fermionic dipolar gases that are confined to two- or three-dimensional lattice potentials \cite{Goral2002qpo,Barnett2006qmw,Menotti2007mso,Buchler2007sc2,Yi2007nqp,Danshita2009sos,Cooper2009sts,Capogrosso2010qpo,Pollet2010spw,Potter2010sad,Pikovski2010isi,Li2010cmo}. In particular, strongly interacting one-dimensional (1D) gases with power-law interactions are expected to become experimentally accessible \cite{Dalmonte2010oqf}.  This would allow us to build on current progress in our group involving strongly interacting 1D systems with contact interactions \cite{Haller2009roa,Haller2010pqp}.

In all quantum gas experiments, the phase-space density is a crucial figure of merit. For atomic samples, most of the gain in phase-space density is typically achieved by laser cooling. Once temperatures are sufficiently low and densities are sufficiently high, evaporative cooling is used to bring the gas to quantum degeneracy. Both of these traditional cooling techniques are hard or even impossible to implement for molecular gases. Molecules usually lack the cycling transitions needed for laser cooling, even though recent experimental progress indicates that this is not always the case \cite{Shuman2010lco}.  Inelastic collisions between the molecules or with coolant gas are additional problems which have thus far prevented molecular gases from being evaporatively cooled. To efficiently create high phase-space density samples of ground-state molecules, we intend to create weakly bound Feshbach molecules \cite{Regal2003cum,Herbig2003poa} out of a quantum gas mixture by means of a Feshbach resonance \cite{Kohler2006poc,Chin2010fri}. The molecules will then be transferred into the rovibronic ground state using the stimulated Raman adiabatic passage (STIRAP) technique \cite{Bergmann1998cpt}. Recently, STIRAP has been used on ultracold and dense samples of Feshbach molecules for efficient two-photon transfer into the rovibronic ground state of the $a \, ^3\Sigma_u^+$ potential for Rb$_2$ molecules \cite{Lang2008utm}, into deeply bound vibrational levels of the $X \, ^1\Sigma_g^+$ potential for Cs$_2$ molecules \cite{Danzl2008qgo}, and into the rovibronic ground state for polar KRb molecules \cite{Ni2008ahp}.  A two-step (four-photon) STIRAP process has been used for transfer into the rovibronic ground state of Cs$_2$ \cite{Danzl2010auh}.  An alternative approach to prepare molecular samples in the rovibronic ground state is based on photoassociation schemes \cite{Sage2005opo,Deiglmayr2008fou,Viteau2008opa,Aikawa2010cto}.

Our goal is to generate a dipolar quantum gas of RbCs. This molecule belongs to a class of dimer molecules chemically stable under two-body collisions \cite{Zuchowski2010rou}, i.e., the reaction RbCs + RbCs $\to$ Rb$_2$ + Cs$_2$ is endothermic when RbCs is in its rovibronic ground state. As a consequence, RbCs dipolar quantum gases are expected to be stable. In contrast, KRb is susceptible to the process KRb + KRb $\to$ K$_2$ + Rb$_2$ \cite{Ospelkaus2010qcc}. In general, we expect that at ultralow temperatures, and in the lowest hyperfine sublevel of the rovibronic ground state of the RbCs molecule \cite{Aldegunde2008hel}, only molecular three-body collisions present a possible limitation on performing experiments with dipolar many-body systems. In the present work, we discuss a crucial step towards the production of a quantum-degenerate RbCs gas: the simultaneous generation of a BEC of Rb and a BEC of Cs atoms. The two BECs are created in separate traps.  After preparation, we can merge them for mixed-species quantum gas experiments and for the generation of RbCs Feshbach molecules as precursors to RbCs ground state molecules.

\section{R\lowercase{b}-C\lowercase{s} mixture in reservoir trap}
\label{sec:1}
The beginning of the experimental cycle is similar to that used for our measurements of interspecies Rb-Cs Feshbach resonances \cite{Pilch2009ooi}. In brief, we load a two-species magneto-optical trap (MOT) from a two-species Zeeman-slowed atomic beam.  The MOTs are separated from each other slightly by careful adjustment of the radiation pressure effect of the Zeeman-slowing laser, allowing us to greatly reduce light-induced collisional losses.  Then the MOTs, each typically containing $1\!\times\!10^{8}$ atoms, are compressed and overlapped.  The atoms are then cooled into a large-volume ``reservoir'' \cite{Kraemer2004opo,Pilch2009ooi} optical dipole force trap, using degenerate Raman-sideband cooling (DRSC) \cite{Kerman2000bom,Treutlein2001hba,Kraemer2004opo}.  DRSC polarizes the atoms in their absolute hyperfine Zeeman ground states: Rb ($F=1, ~ m_F=1$) and Cs ($F=3, ~ m_F=3$).  The reservoir trap serves as a source of atoms for loading tighter ``dimple'' dipole traps \cite{Kraemer2004opo}.  The high curvature of the dimples makes them ideal for building up phase-space density through evaporation.  All trapping beams (reservoir and dimples) are turned on 4~s before DRSC.  The dimples are loaded by slowly ramping down the reservoir trap power.

Compared to our previous work \cite{Pilch2009ooi}, we were able to improve the DRSC cooling stage in three ways \cite{Lercher2010}: First, the lattice light for trapping and cooling the Rb atoms is now about 10~GHz blue-detuned from the $F=1 \to F'=2$ component of the $^{87}$Rb D2 line. Second, we have added a depumping beam for the Rb atoms, which is resonant with the $F=2 \to F'=2$ transition. This allows us to shorten the DRSC pulse to 3 ms. And third, we now apply two or three subsequent pulses of DRSC for both species\footnote{The application of several pulses of DRSC increases the number of atoms loaded into the dimple traps. The time between the pulses is approximately one fourth of the reservoir trap oscillation period.  Atoms cooled by a DRSC pulse at the edge of the reservoir trap gain kinetic energy afterwards as they fall towards the center.  Subsequent DRSC pulses remove some of this kinetic energy.}. When no dipole traps are present, we measure about $3\!\times\!10^{7}$ atoms and temperatures of about $1.5~\mu$K for both species when we adiabatically release the atoms into free space.

\begin{figure}[htbp]
 \includegraphics[width=8.5cm] {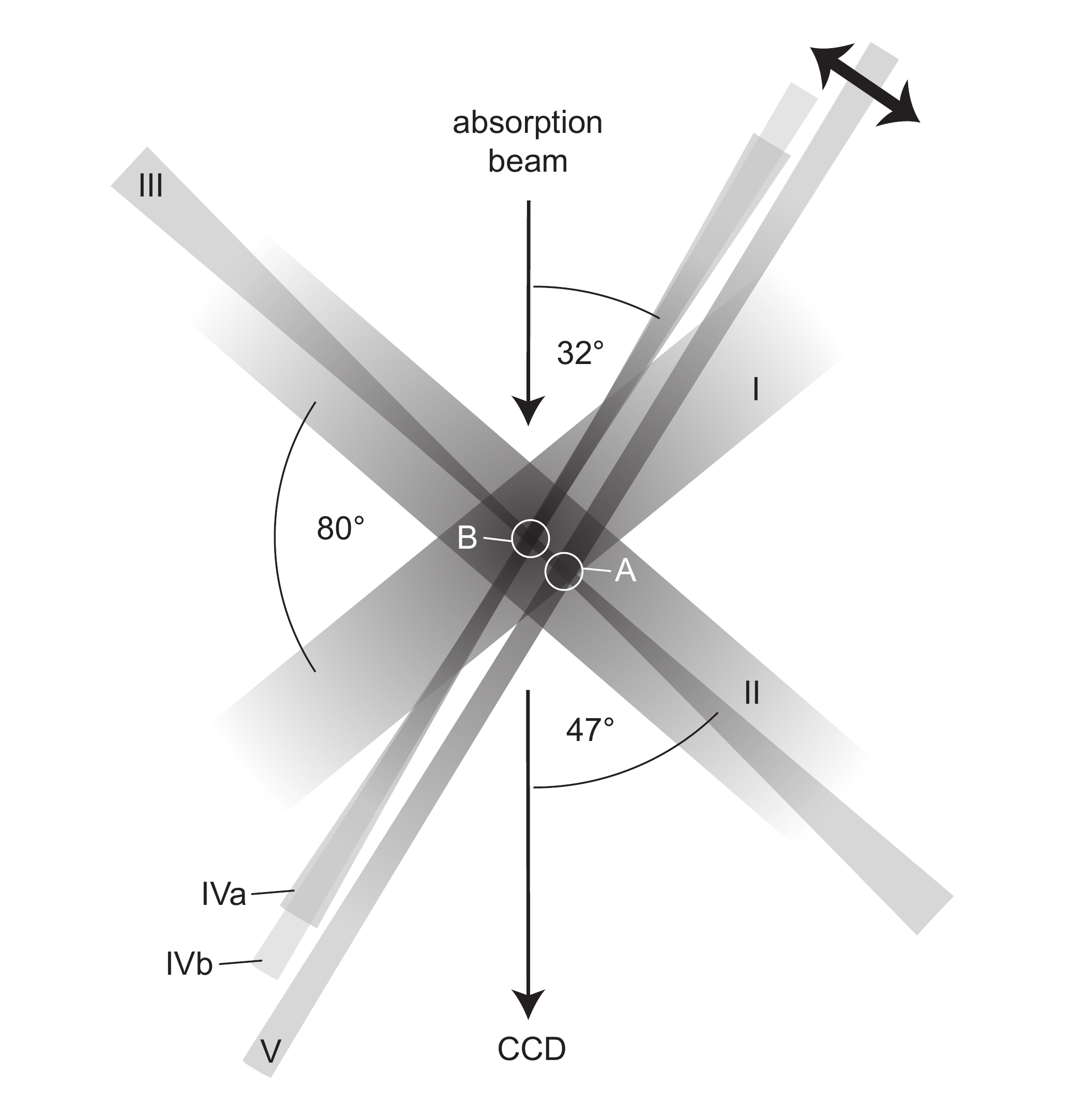}
  \caption{Trap configuration. All laser beams lie in the same horizontal plane. Beams I and II form the large-volume reservoir trap. Co-propagating beams IVa and IVb cross beam III to form the two-color dimple trap for Rb (trap B). Beams V and III form the dimple trap for Cs (trap A). Beams II and III are almost, but not quite collinear.  Translation of beam V allows the two dimples to be merged, as indicated by the arrow.  Absorption imaging is performed at the angles indicated.
  }
  \label{beams}
\end{figure}

The reservoir trap (beams I and II in Fig.~\ref{beams}) uses broadband light at 1070~nm in two crossed, horizontal laser beams at a power of 43.5~W each. The Rb atoms experience a confining potential that is a factor of 1.7 shallower than the one for Cs as a result of the lower optical polarizability of Rb at this wavelength. We apply a vertical magnetic field gradient of 30.7~G/cm to hold the atoms against gravity. The optimal levitation gradients for Rb ($F=1, ~ m_F=1$) and Cs ($F=3, ~ m_F=3$) are nearly identical due to a fortuitous coincidence in the magnetic-moment-to-mass ratios.  With the dimple beams disabled, we can investigate the properties of the reservoir alone.  We find that there is rapid thermalization of the two species.  After holding the reservoir trap depth constant (plain evaporation) for 900~ms after DRSC, we measure a common temperature of about $5~\mu$K.  There are $1.7\!\times\!10^{6}$ atoms of each species, with a peak density of $\!1\times 10^{11}$~cm$^{-3}$. At these densities, intra- and inter-species three-body loss is still negligible. The peak phase-space density is about $3\!\times\!10^{-5}$ for both species.

\begin{table}[htbp]
\caption{Initial trap parameters.  The waist radius given here is the $1/e^{2}$ intensity radius.}
\label{beamparams}
\begin{tabular}{l|lllll}
\hline\noalign{\smallskip}
Beam&I,II&III&IVa&IVb&V\\
\noalign{\smallskip}\hline\noalign{\smallskip}
Wavelength (nm)&1070&1064.5&820&1064.5&1064.5 \\
Waist radius ($\mu$m)&500&45&66&57&102\\
Rb depth / $k_B$ ($\mu$K)&13&8.5&24&10&6.0\\
Rb rad. freq. (Hz)&23&204&234&176&77\\
Cs depth / $k_B$ ($\mu$K)&24&15&-25&18&11\\
Cs rad. freq. (Hz)&25&220&192i&189&83\\
\noalign{\smallskip}\hline
\end{tabular}
\end{table}

\section{Condensation of R\lowercase{b} and C\lowercase{s}}
\label{sec:2}
Our initial attempts to evaporate Rb and Cs together in the same dimple were plagued by high interspecies collisional losses.  We therefore decided to pursue a strategy in which we separate the two species before forced evaporation.  The laser beam arrangement used to perform these experiments is shown in Fig.~\ref{beams}. The wide I and II beams form the reservoir trap as discussed in the previous section. Beams III-V form two dimple traps.  The dimple trap for Cs (trap A in Fig.~\ref{beams}) lies at the intersection of beams III and V, both at a wavelength of $1064.5$~nm.  The dimple trap for Rb (trap B in Fig.~\ref{beams}) results from crossing co-propagating beams IVa and IVb with beam III. Beam IVa at 820~nm is blue-detuned with respect to the Cs D1 and D2 transitions, but red-detuned from the Rb transitions.  Beam IVb is also at 1064.5~nm. Because all of the $1064.5$~nm beams come from the same single-frequency laser, the acousto-optic modulators used to regulate their powers are configured to give crossing beam pairs different frequencies to avoid standing waves. The frequency of beam IVa is tuned to a large gap between Rb-Rb photoassociation resonances. Table~\ref{beamparams} summarizes the trap properties at maximum power (before ramping down).

Under high-density loading conditions, dimple trap A contains only Cs after the reservoir is ramped down.  The high Rb-Cs thermalization rate gives the two species similar temperatures.  However, the Cs trap depth is 1.7 times that for Rb.  This means that the truncation parameters $\eta$~=  trap depth / ($k_B\times$ temperature), and thus the evaporation rates of the two species differ greatly.  Rb acts essentially as a coolant for Cs here.  In dimple trap B, we can adjust the relative powers of beams IVa and IVb to invert the Rb / Cs trap depth ratio compared to that in trap A, and we load only Rb here.  Note that beam IVa is only needed during the loading process for trap B and the first evaporation ramp. Afterwards, we shut it off to reduce Rb depolarization.  The inverted trap depth ratios for the two dimples causes the heat load to be distributed more evenly between Rb and Cs in the reservoir.  In fact, we load more Cs atoms into trap A when the Rb atoms are present in the reservoir and vice versa. For control of the location of trap A, beam V can be steered by a piezoelectric transducer.  The traps can thus be overlapped for collisional studies (Sect.~\ref{sec:3}) and to create heteronuclear molecules \cite{Takekoshi2010b}.

While trap B lies at the center of the reservoir trap, where the Rb density and hence the Rb-Rb and Rb-Cs collision rates are highest, trap A is initially offset by 130 $\mu$m to avoid Cs-Cs-Cs three-body loss.  We control the Cs-Cs scattering length by tuning the magnetic field \cite{Kraemer2004opo}, starting at $660~a_\mathrm{o}$ ($34.1$~G) in order to keep the Cs-Cs collision rate high in the reservoir.  After DRSC we wait 900 ms (plain evaporation), and then ramp the reservoir power down to zero in $1500$~ms.  At this point, the dimples typically contain $2.6\!\times\!10^{5}$ Cs atoms and $3.6\!\times\!10^{5}$ Rb atoms at temperatures of $2.8~\mu$K and $2.4~\mu$K, respectively. The peak densities are about $1\!\times\!10^{13}$~cm$^{-3}$, and the phase-space densities have increased by three orders of magnitude with respect to their values in the reservoir trap.  The elastic scattering rates are $\approx\!200$~s$^{-1}$ for Rb and $\approx\!500$~s$^{-1}$ for Cs, and we thus have excellent starting conditions for evaporative cooling.

We now start a sequence of three exponential forced evaporation ramps. The total duration of these ramps is $9-12$ s. We follow the evaporation procedure used for generating a BEC of Cs atoms \cite{Kraemer2004opo}. During these ramps, the Cs-Cs scattering length is magnetically tuned to give optimal elastic scattering rates while minimizing three-body loss.  The quadrupole magnetic field used to levitate the atoms also creates a broad radial antitrapping potential which cancels the trapping potentials of each dimple beam in the axial direction.  The effective trap depth thus becomes that of the weaker dimple beam in the crossed pair. We find, therefore, that it is important that the atoms leave their respective traps along beams IVa,b and V.  Atoms leaving along beam III give rise to losses when they enter the other trap. To prevent this, we always keep the trap depth along the direction of beam III higher than that in the direction of beams IVa,b and V. We thus primarily lower the power in beam III.

At the beginning of the first ramp, we increase the trap separation to 420~$\mu$m to reduce residual atom exchange between the traps. The first ramp lasts 250~ms, and we lower the power in beam III to 35\% of the initial power of 210~mW. The duration of the second ramp is 500~ms. Here, we lower the power to 30\% of the initial power. At the end of this ramp, the Rb cloud is still thermal, but is close to the BEC phase transition.

\begin{figure*}[htpb]
 \includegraphics[width=6.5in] {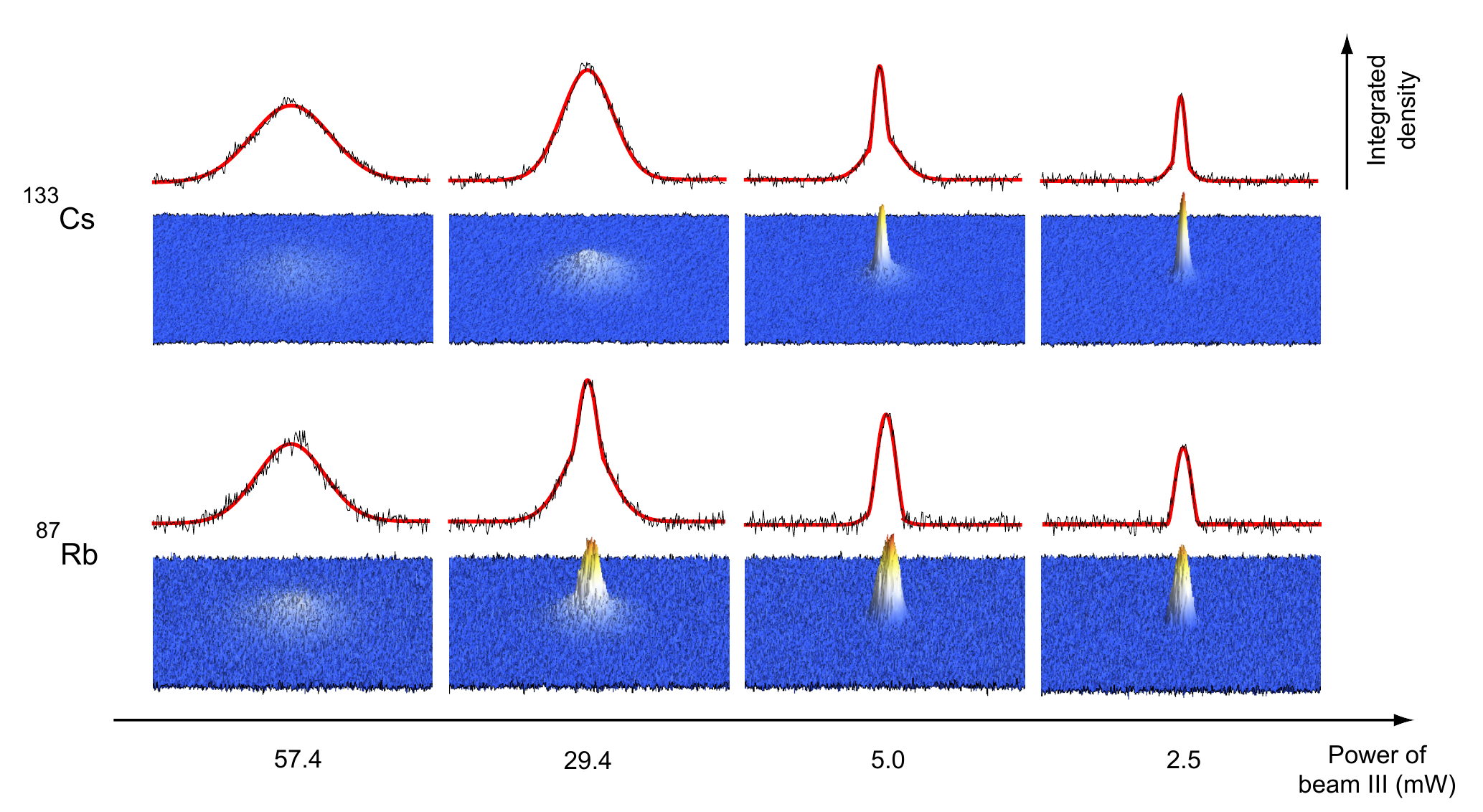}
  \caption{(color online) Simultaneous BEC formation for Cs (top row) and Rb (bottom row) in separated traps.  Absorption images after $25$ ms of expansion at a magnetic field of $17.4$~G, are taken at different times during the final evaporation ramp. The power in beam III is indicated. The fits (solid lines) to the integrated density profiles give clear evidence of the phase transition as they change from Gaussian to bimodal and finally to a Thomas-Fermi distribution.  For the first three pictures (from left to right) the final ramp duration is 8~s, and for the last picture 11~s.  
\label{condensates}}
\end{figure*}

We observe the onset of condensation in both traps during the third evaporation ramp.  This occurs approximately at the calculated transition temperatures: $260~$nK for Rb and $180~$nK for Cs.  The transition from a Gaussian, to a bimodal, and finally to a parabolic integrated density profile can be clearly seen in Fig.~\ref{condensates}.  The third ramp lasts $8-11$~s, bringing the power to the values indicated in Fig.~\ref{condensates}.  Both BECs have a $1/e$ lifetime of about 8~s, consistent with loss due to collisions with background gas. In Table \ref{tab_beceigenschaften} we list the relevant parameters for both BECs at the end of the last evaporation ramp.  We are not able to extract any Rb thermal background from the fits, but the Cs cloud has a small ($<10\%$) thermal component.

At this point, our dual-species BEC has particle numbers and trapping parameters similar to those in the $^{85}$Rb-$^{87}$Rb BEC at JILA \cite{Papp2008tmi} and the $^{41}$K-$^{87}$Rb BEC realized by the LENS group \cite{Modugno2002tas,Catani2008dbm,Thalhammer2008dsb}. It represents a good starting point for planned experiments exploring dual-species SF-MI physics, the creation of molecular quantum gases, condensate miscibility, and the study of Efimov physics in condensate collisions \cite{Wang2010cbe}. We expect that, with some optimization of the cooling and trapping procedure, we will be able to produce BECs each containing $10^5$ particles or more. While we believe that atom exchange between the dimples during the evaporation sequence is not a limiting factor, the fact that we do not want to evaporate along the shared dimple beam (beam III) reduces our efficiency.  One solution would be to fully decouple both samples by introducing a second trapping beam along the direction of beam III, but offset vertically.  Subsequent experiments using underlevitation to evaporate downwards increased the size of our condensates, but will not be presented here. Finally, since beam IVa at 820 nm depolarizes the Rb as a result of an initial photon scattering rate of about 1~s$^{-1}$, it might be advantageous to tune its wavelength further to the red, if enough power is available.

\begin{table}[htpb]
\caption{Rb-Cs dual-species BEC parameters.  Atom numbers and trap frequencies are measured.  The Thomas-Fermi radii, chemical potentials, and peak densities are calculated.  The combined statistical and systematic error is 10 to 20~\% for the atom number and 1 to 2~\% for the trap frequencies.  The Rb-Rb scattering length is taken from Ref. \cite{Strauss2010hrv}.  The Cs-Cs scattering length is estimated using Ref. \cite{Lange2009doa}.}
\label{tab_beceigenschaften}
\begin{tabular}{l|ll}
\hline\noalign{\smallskip}
  & $^{87}$Rb BEC & $^{133}$Cs BEC\\
\noalign{\smallskip}\hline\noalign{\smallskip}
Atom number  & $2.0\times 10^{4}$ & $1.0\times 10^{4}$ \\
Scattering length~($\mathrm{a}_0$)&$100.4$&$189$ \\
Radial trap freq. (Hz) &(III) 22&(III) 24\\
  & (IVb) 124& (V) $44$\\
Thomas-Fermi radius ($\mu$m)&(III) 17& (III) 9.0\\
  & (IVb) 3.0&(V) 4.9\\
Chem. potential / $k_B$ (nK) &290&150\\
Peak density (cm$^{-3}$)&$7.8\times10^{13}$ & $3.2\times10^{13}$\\
\hline\noalign{\smallskip}
\end{tabular}
\end{table}

\section{Interspecies collisional loss}
\label{sec:3}
Because rapid thermalization indicates a large interspecies background scattering length, one expects to encounter a high three-body recombination rate in Rb-Cs mixtures. In fact, magnetic field modulation spectroscopy shows that there is a weakly bound Feshbach level with a binding energy of $h \times 115(20)$ kHz \cite{Takekoshi2010}.  This corresponds\footnote{To calculate the interspecies background scattering length, we assume the mean scattering length to be $86.24~\mathrm{a}_0$. This comes from the Van der Waals coefficient $C_{6}$ given in \cite{Derevianko2001hpc}.} to a Rb-Cs scattering length of $+630(60)~\mathrm{a}_0$.  In order to understand these losses, we have performed measurements on an ultracold Rb-Cs mixture created by overlapping the two dimples after evaporation.  Similar measurements have been performed using K-Rb mixtures evaporatively cooled together in a single trap \cite{Ospelkaus2006idd,Barontini2009ooh}.

We use high phase-space density thermal clouds here because the condensates were found to be immiscible.  Both trap depths are ramped down as the dimples are combined.  The final trap depths are low enough ($k_B \times$ 510nK  for Rb and $k_B \times$ 880nK for Cs) so that all three-body recombination products can escape.  The starting peak densities in the mixture are $6\times 10^{12}$ cm$^{-3}$ for Rb and $7\times 10^{12}$ cm$^{-3}$ for Cs.  In order to reduce intraspecies collisions, the magnetic field is set near 17~G so that the Cs-Cs scattering length is zero.  The peak elastic scattering rate for Rb starts at $10~$s$^{-1}$, which is much smaller than the Rb and Cs trap frequencies.  To observe the one-body losses due to collisions with background gas, one species can be removed by radiation pressure at the beginning of the thermalization period.

\begin{figure}[t]
 \includegraphics[width=8.5cm] {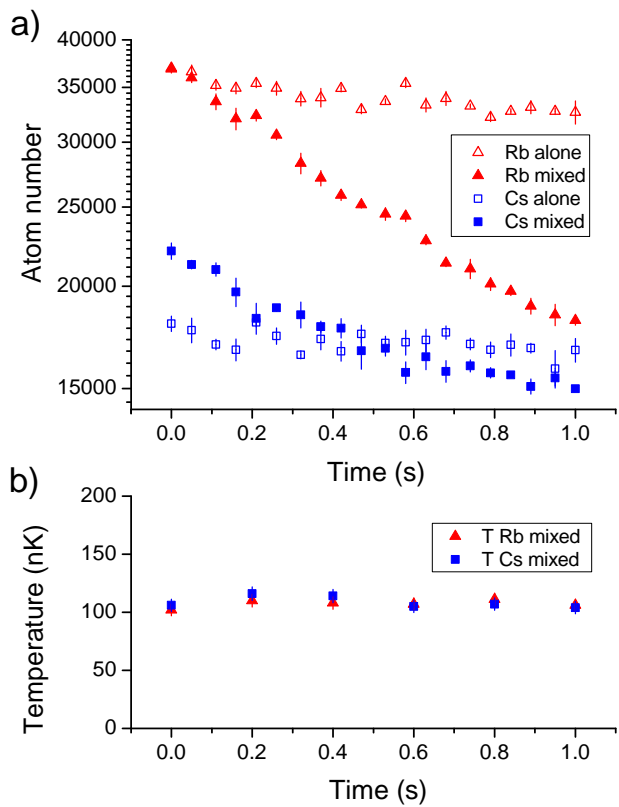}
 \caption{Trap loss measurements for an uncondensed high-density Rb-Cs mixture: a) atom numbers (logarithmic scale) and b) temperatures as a function of hold time.  Triangles: Rb, squares: Cs, as a function of hold time. Open symbols: alone in trap.  Closed symbols: in mixture.  The effective trap depth is $k_B \times$ 510nK  for Rb and $k_B \times$ 880nK for Cs.
 \label{mixture}}
\end{figure}

The results, shown in Fig. \ref{mixture}, are typical for high density Rb-Cs mixtures over a wide range of trap depths.  Compared to each species alone, there is a large additional loss due to interspecies collisions.  In such a mixture, the truncation parameters remain anomalously low ($\eta<10$).  For example, if the trap depths are doubled by adiabatically doubling the powers of both dimple beams, the resulting curves are almost identical to those in Fig. \ref{mixture}, but the temperature also doubles.  In the absence of heating, this temperature should scale as (trap depth)$^{1/2}$.  Evaporation cannot be shut off while the density is high.  In the mixture shown, the temperature actually remains constant during the loss.  We interpret this as a balance between a heating effect (most likely three-body recombination) and evaporation.

It is also possible to verify that three-body loss alone is not enough to account for all of the observed loss.  Extraction of the three-body loss coefficients from these curves is difficult and requires a detailed model of the evaporation.  However, a naive fit using the one-body loss constants observed for Rb and Cs when alone gives three-body trap loss coefficients on the order of $10^{-25}$ cm$^6$s$^{-1}$.  This is much larger than the coefficients expected based on the known two-body scattering lengths \cite{Braaten2007epi}.  We therefore assume that a significant portion of the observed loss is due to evaporation.  Two-body loss due to absorption of trapping laser photons is negligible because no additional losses are observed when the laser powers are doubled.

An obvious way to reduce these losses would be to find a magnetic field at which the Rb-Rb, Rb-Cs, and Cs-Cs scattering lengths are all small, but we have confirmed that no such region exists in the range of fields available to us ($<667$~G) \cite{Takekoshi2010}.  The timescale of these losses, however, is longer than that required for magnetoassociation, and we find that we can create Feshbach molecules using these mixtures at efficiencies approaching 5\% \cite{Takekoshi2010b}.

Our method of combining separately prepared BECs is a general technique to study high-density mixtures of ``problem'' atoms with interspecies collisional properties unfavorable for evaporation.

\section{Towards a quantum gas of R\lowercase{b}C\lowercase{s} molecules}
\label{sec:4}
To optimize the overall molecule production efficiency by minimizing collisional loss during the preparation procedure, which is particularly crucial for bosonic molecules such as RbCs, we plan to perform Feshbach association and ground state transfer in the presence of a three-dimensional optical lattice with precisely one atom pair per lattice site \cite{Jaksch2002coa,Moore2003ctm,Freericks2010ite}. This has been recently demonstrated for the simpler case of Cs$_2$ \cite{Danzl2010auh}.  A suitable wavelength for the lattice light is 1064.5 nm.  This is easily produced using high-power fiber amplifiers.  The ac polarizability of RbCs Feshbach molecules and RbCs rovibronic ground-state molecules happens to be almost identical at this wavelength \cite{Dulieu2010} (just as in the case of Cs$_2$ \cite{Danzl2010auh,Vexiau2011}).

The central question is how to generate a many-body state in the lattice for which the number of Rb-Cs pairs as precursors to RbCs molecules is maximized. In view of considerable Cs-Cs-Cs, Rb-Rb-Cs and Rb-Cs-Cs three-body loss, we will load the lattice while the two samples are still separated.  Before raising the lattice depth, the two clouds will be brought as close as possible to each other, so that they are both covered by the lattice. This is easily achieved by moving the laser beams forming the two dimple traps.  While the two atomic samples are still separated, we are free to choose the magnetic field. For concreteness, we will choose $21.0$~G, for which the Cs-Cs scattering length is $210~\mathrm{a}_0$, to minimize Cs-Cs-Cs three-body loss \cite{Kraemer2006efe}.  When raising the lattice depth we will, in a species-selective way, drive the SF-MI phase transition: the Cs sample will undergo the transition at a Cs lattice depth of $\approx 12~E_\mathrm{R}^\mathrm{Cs}$, where $E_\mathrm{R}^\mathrm{Cs}$ is the recoil energy for the Cs atoms.  The external dipole trap confinement at the position of the Cs atoms will be adjusted in such a way that the occupation of the one-atom shell of the MI state is maximized, while the Cs lattice depth is raised to $\approx 20~E_\mathrm{R}^\mathrm{Cs}$ to prevent the Cs atoms from tunneling.  At this lattice power, the Rb sample is still deeply in the superfluid regime. It can then be steered onto the Cs sample by merging the dimple traps.  Cs-Cs-Cs and Rb-Cs-Cs three-body loss should be fully quenched, while Rb-Rb-Cs three-body loss can be suppressed by making the kinetic energy associated with Rb tunneling much smaller than the Rb-Rb interaction energy. A further increase of the lattice power together with control of the Rb-Cs scattering length should then also freeze out the Rb sample in such a way that precisely one Rb atom pairs up with one of the Cs atoms.  Note that weak attractions might enhance the pair creation probability at the edge of the one-atom shell, where the Cs atoms do not occupy every lattice site \cite{Freericks2010}.

We are aware of related work on the production of a Rb-Cs BEC by S. Cornish's group at Durham University, UK \cite{Cornish2011}.

\begin{acknowledgement}
We thank D. Baier for contributions at an early stage of the experiment. We acknowledge support by the Austrian Science Fund (FWF) and the European Science Foundation (ESF) within the EuroQUAM / QuDipMol project (FWF project number I124-N16) and support by the FWF through the SFB FoQuS (FWF project number F4006-N16).
\end{acknowledgement}


\end{document}